\documentstyle[12pt]{article}
\def\ie{{\em i.e.}}
\def\eg{{\em e.g.}}

\def\beq{\begin{equation}}
\def\eeq{\end{equation}}

\catcode`\@=11 
\def\coeff#1#2{{\textstyle{#1\over #2}}}

\def\VEV#1{\left\langle #1\right\rangle}

\def\lsim{\mathrel{\mathpalette\@versim<}}
\def\gsim{\mathrel{\mathpalette\@versim>}}
\def\@versim#1#2{\vcenter{\offinterlineskip
    \ialign{$\m@th#1\hfil##\hfil$\crcr#2\crcr\sim\crcr } }}

\def\JL{J. L. Lopez}
\def\DVN{D. V. Nanopoulos}

\def\t1{{\tilde 1}}

\def\GeV{\,{\rm GeV}}
\def\TeV{\,{\rm TeV}}

\def\to{\rightarrow}

\def\a{(U+\bar U)}
\def\b{(T+\bar T)}
\def\c{(\phi+\bar\phi)}

\def\h{\partial_\phi\ln W}
\def\hb{\partial_{\bar\phi}\ln\bar W}
\def\NPB#1#2#3{Nucl. Phys. B {\bf#1} (19#2) #3}
\def\PLB#1#2#3{Phys. Lett. B {\bf#1} (19#2) #3}

\def\PRD#1#2#3{Phys. Rev. D {\bf#1} (19#2) #3}

\begin{document}
\begin{flushright}
\baselineskip=12pt
{CERN-TH.7433/94}\\
{CTP-TAMU-40/94}\\
{ACT-13/94}\\
{MIU-THP-69/94}\\
{hep-ph/9409223}
\end{flushright}

\begin{center}
{\Huge\bf New phenomena in the standard no-scale supergravity model\\}
\vglue 0.25cm
{S. KELLEY$^{(b),(e)}$, JORGE L. LOPEZ$^{(a),(b)}$, and D.V.
NANOPOULOS$^{(a),(b),(c)}$,\\ and A. ZICHICHI$^{(d)}$\\}
\vglue 0.2cm
{\em $^{(a)}$Center for Theoretical Physics, Department of Physics, Texas A\&M
University\\}
{\em College Station, TX 77843--4242, USA\\}
{\em $^{(b)}$Astroparticle Physics Group, Houston Advanced Research Center
(HARC)\\}
{\em The Mitchell Campus, The Woodlands, TX 77381, USA\\}
{\em $^{(c)}$CERN Theory Division, 1211 Geneva 23, Switzerland\\}
{\em $^{(d)}$CERN, 1211 Geneva 23, Switzerland\\}
{\em $^{(e)}$Deparment of Physics, Maharishi International University\\
Fairfield, Iowa~52557--1069\\}
\baselineskip=12pt

\vglue 0.25cm
ABSTRACT
\end{center}
{\rightskip=3pc
 \leftskip=3pc
\noindent
\baselineskip=20pt
We revisit the no-scale mechanism in the context of the simplest no-scale
supergravity extension of the Standard Model. This model has the usual
five-dimensional parameter space plus an additional parameter $\xi_{3/2}\equiv
m_{3/2}/m_{1/2}$.  We show how predictions of the model may be extracted over
the whole parameter space. A necessary condition for the potential to be stable
is  ${\rm Str}{\cal M}^4>0$, which is satisfied if $\bf m_{3/2}\lsim2 m_{\tilde
q}$. Order of magnitude calculations reveal a no-lose theorem guaranteeing
interesting and potentially observable new phenomena in the neutral scalar
sector of the theory which would constitute a ``smoking gun'' of the no-scale
mechanism. This new phenomenology is model-independent and divides into three
scenarios, depending on the ratio of the weak scale to the vev at the minimum
of the no-scale direction. We also calculate the residual vacuum energy at the
unification scale ($C_0\, m^4_{3/2}$), and find that in typical models one must
require $C_0>10$. Such constraints should be important in the search for the
correct string no-scale supergravity model. We also show how specific classes
of string models fit within this framework.}


\begin{flushleft}
\baselineskip=12pt
{CERN-TH.7433/94}\\
{CTP-TAMU-40/94}\\
{ACT-13/94}\\
{MIU-THP-69/94}\\
August 1994
\end{flushleft}

\vfill\eject
\setcounter{page}{1}
\pagestyle{plain}
\baselineskip=14pt

\section{Introduction}
The search for a unified theory of everything has two boundaries. At one
extreme the Standard Model matches flawlessly with experiment, and at the other
extreme string theory promises quantum consistent unification of gravity with
the fields of lower spin.  In connecting these extremes, two vital clues have
inspired major progress over the years: the gauge hierarchy problem, and the
problem of vanishing cosmological constant. Global supersymmetry provides a
first solution to the gauge hierarchy
problem and ensures the cancellation of $\Lambda^4$ divergent one-loop
contributions to the vacuum energy. However, global supersymmetry is not
enough: we ultimately need local supersymmetry --- supergravity. In addition,
to simultaneously preserve the gauge hierarchy and respect the present
experimental non-observation of sparticles, supergravity must be broken to a
globally supersymmetric theory with soft breaking terms of order of the
electroweak scale.

Supergravity is described in terms of two functions: the K\"ahler
function ($G$) and the gauge kinetic function ($f_{\alpha\beta}$), plus the
gauge and (hidden plus observable) matter content of the theory. Furthermore,
the supersymmetry breaking scale ($\widetilde m$) is not assured to be
comparable to the electroweak scale. Moreover, in the process a large vacuum
energy is usually generated (${\cal O}(\widetilde m^2M^2_{Pl})={\cal
O}(10^{-32}M^4_{Pl})$). A very important exception to this typical and
unsatisfactory situation occurs in a class of supergravity models with a
distinct K\"ahler function endowed
with some non-compact symmetries (\eg, $SU(1,1)$ or $SU(N,1)$)
\cite{Cremmer,EKNI+II,LN}. This class of theories have three remarkable
properties after supersymmetry breaking: (i) the minimum of the scalar
potential is at zero vacuum energy \cite{Cremmer}, (ii) at the minimum there is
one or more flat directions (moduli) which leave the gravitino mass
undetermined \cite{Cremmer,EKNI+II}, and (iii) there are no large one-loop
corrections to the vacuum energy ($\propto {\rm Str}{\cal M}^2$) which would
destroy the flat directions \cite{EKNI+II}. The first property implies that the
vacuum energy is ${\cal O}(M^4_Z)={\cal O}(10^{-64}M^4_{Pl})$ or smaller. The
second property provides a natural solution to the gauge hierarchy problem via
the {\em no-scale} mechanism \cite{Lahanas,EKNI+II}, whereby the minimization
of the electroweak potential determines the vacuum expectation value of the
Higgs fields and of the moduli fields associated with the flat directions, thus
determining the scale of supersymmetry breaking to be comparable to the
electroweak scale. The third property ensures that the no-scale mechanism is
stable under radiative corrections which would otherwise jeopardize the gauge
hierarchy.

The above class of models were first proposed in the context of supergravity
{\em per se}, and have been since found to be strongly supported by the
low-energy effective theories from string
\cite{Witten,FGKP,FKPZI,FKPZII,PZ,LNY94,FKZ}. In fact, in string model-building
the class of ``{\em string no-scale supergravities}" is much extended
\cite{FKPZII,PZ} to include new kinds of moduli fields and more definite forms
for the K\"ahler function \cite{FGKP,LNY94}. Moreover, in string models one can
accommodate the usual non-perturbative (\eg, gaugino condensation)
supersymmetry breaking mechanism, as well as tree-level breaking via
coordinate-dependent compactifications \cite{FKPZII,PZ}. All these features
make string no-scale supergravity a very rich, interesting, and well motivated
subset of all possible string supergravities. Furthermore, the three properties
mentioned above are expected to be very discriminating in the selection of
phenomenologically appealing string vacua.

In this paper we explicitly solve the minimal no-scale supergravity extension
of the Standard Model.  We identify the minimal parameter space of this model
and determine the other parameters by minimization with respect to the Higgs
fields and the additional ``{\em no-scale}" {\em direction}. A necessary
condition for stability of the potential in the no-scale direction is ${\rm
Str}{\cal M}^4>0$, which requires $m_{3/2}\lsim2m_{\tilde q}$.  The real and
imaginary degrees of freedom corresponding to the additional {\em no-scale
field} lead to new phenomenology in the neutral scalar sector. Calculation of
the residual vacuum energy at the unification scale as a function of the
soft-supersymmetry-breaking parameters leads to interesting constraints on the
parameters of the model.

\section{The standard no-scale supergravity model}
\label{sec:standard}
Consider the simplest supergravity extension of the Standard Model.  Numerous
previous studies have identified the parameter space of this model, shown how
to solve the model, and extracted its predictions as a function of this
five-dimensional parameter space which can be taken as
($\tan\beta$,$m_t$,$m_{1/2}$,$\xi_0$,$\xi_A)$, with $\xi_0\equiv m_0/m_{1/2}$
and $\xi_A\equiv A/m_{1/2}$. However, the naive construction of the
one-loop effective potential \cite{KLNQ+GRZ} has a major flaw: it is not
formally independent of the renormalization scale, although derivatives with
respect to the various Higgs fields are. This problem was studied in
Ref.~\cite{aspects}, where a simple and well motivated ansatz for a
$Q$-independent one-loop effective potential was proposed: one should subtract
the field-independent contribution to the potential. That is, one should
use the following expression
\begin{equation}
V_1=V_0+{1\over64\pi^2}{\rm Str}\,{\cal M}^4\left(\ln{{\cal
M}^2\over Q^2}-{3\over2}\right)-V_1(0),
\label{eq:V1}
\end{equation}
where $V_0$ is the usual RGE-improved tree-level Higgs potential, $V_1(0)$
is the field-independent contribution to $V_1$, and
the supertrace (${\rm Str}\,{\cal M}^n=\sum_j(-1)^{2j}(2j+1){\rm Tr}\,{\cal
M}^n_j$) includes a term for the gravitino. Although the field-independent term
and the contribution of the gravitino in the supertrace are irrelevant to
minimization with respect to the Higgs fields, these two terms are crucial to
dynamically determining supersymmetry breaking.

Since the field-independent contribution to $V_1$ could contain an unknown
$Q$-independent piece, we parametrize it more generally as $C\,m^4_{3/2}$. The
RGE satisfied by $C$ is easily derived by demanding that $V_1$ be
scale-independent to one-loop order (see Eq.~(2.15) of Ref.~\cite{aspects})
\begin{equation}
{dV_1\over dt}={dV_0\over dt}-{1\over32\pi^2}{\rm
Str}{\cal M}^4+m^4_{3/2}{dC\over dt}+({\rm ``two-loop"})=0.
\end{equation}
Moreover, the resulting relation must hold for all values of the fields.
Taking the Higgs fields ($h_1,h_2$) to zero gives $V_0=dV_0/dt=0$, and the RGE
follows
\begin{equation}
m^4_{3/2}{dC\over dt}={1\over32\pi^2}\,[{\rm Str}{\cal M}^4]_{h_{1,2}=0}\ ,
\label{eq:etaRGE}
\end{equation}
with (in the notation of Ref.~\cite{LN})
\begin{eqnarray}
[{\rm Str}{\cal M}^4]_{h_{1,2}=0}&=&
 4 ( m_1^4 + m_2^4 + 2 m_3^4) + 6 ( m_{U^c_3}^4 + m_{D^c_3}^4 + 2 m_{Q_3}^4 )
+ 12 ( m_{U^c}^4 + m_{D^c}^4\nonumber\\
&&+ 2 m_Q^4 )
+ 2 ( m_{E^c_3}^4 + 2 m_{L_3}^4 ) + 4 ( m_{E^c}^4 + 2 m_L^4 ) - 16
M_3^4 - 6 M_2^4 - 2 M_1^4\nonumber\\
&& - 8 \mu^4 -4m^4_{3/2}\ .
\end{eqnarray}

Interestingly enough, there has been a recent effort to identify the value
of $C$ at the scale of supersymmetry breaking ($C_0$) as the remnant
vacuum energy after supersymmetry breaking \cite{KPZ}. Moreover, $C_0$ is in
principle calculable in specific string models and in specific supersymmetry
breaking mechanisms. For example, in tree-level supersymmetry breaking via
coordinate-dependent compactifications one has $C_0\sim(n_B-n_F)$, where
$n_{B(F)}$ is the number of massless bosonic (fermionic) degrees of freedom
after supersymmetry breaking \cite{A,FKZ}.

The essence of the no-scale approach is that the gravitino mass is a function
of the real part of an additional scalar field,  $T_R\equiv {\rm Re}(T)$, whose
vev is undetermined at tree level, \ie, there is a flat direction. We assume
the following functional form
\begin{equation}
m^2_{3/2}={\alpha\Lambda^{2+p}\over{(T_R)^p}}\ ,
\label{eq:m3/2T}
\end{equation}
though it is interesting to consider departures from this. Here $\alpha$ is
a dimensionless parameter and $\Lambda$ is some appropriate mass scale.
It is also possible to have $m_{3/2}$ depend on multiple flat directions, a
possibility which is realized in many string examples. The gravitino mass is
dynamically  determined by the minimum of the loop-corrected potential with
respect to $T_R$.  In this section we derive some model-independent results by
virtue of the fact that the zeroes of the first derivative of the potential
with respect to $T_R$, and the sign of the second derivative are independent of
$\alpha$ and $p$.

To implement the no-scale mechanism one can take one of two approaches:
(i) a top-down approach, were $C_0$ is ``given" and minimization of $V_1$
with respect to $m_{3/2}$ (\ie, the no-scale mechanism) gives $m_{3/2}$; or
(ii) a bottom-up approach, where one uses the no-scale mechanism to determine
$C(M_Z)$ for a given value of $m_{3/2}$ and then obtains $C_0$ by RGE
evolution. Since $C_0$ is in practice rather unknown, here we follow the
bottom-up approach with the hope of finding constraints on the calculated value
of $C_0$ (as a function of the usual soft-supersymmetry-breaking
parameters), which should help guide string model builders in their quest for
models with phenomenologically acceptable values of $C_0$.

We now calculate the first and second derivatives of $V_1$ with respect to
$T_R$. To proceed we scale out the $m_{3/2}$ dependence in $V_1$ (here $X\equiv
m^2_{3/2}$)
\begin{equation}
V_1=X^2\,\widehat V_0 + X^2\,C+{X^2\over64\pi^2}{\rm Str}\widehat{\cal M}^4
\left(\ln{X\widehat{\cal M}^2\over Q^2}-{3\over2}\right),
\end{equation}
where $\widehat V_0=V_0/X^2$, etc. Thus we obtain (see also Ref.~\cite{KPZ})
\begin{equation}
{\partial V_1\over\partial T_R}=-{p\over T_R}X{\partial V_1\over\partial
X}=-{p\over T_R}\left\{2V_1+{1\over64\pi^2}{\rm Str}{\cal M}^4\right\},
\end{equation}
and the no-scale condition ${\partial V_1\over\partial T_R}=0$ then
implies
\begin{equation}
V_1=-{1\over128\pi^2}{\rm Str}{\cal M}^4,
\label{eq:V1min}
\end{equation}
which allows one to determine $C=C(m_{3/2})$ in the bottom-up
approach.\footnote{Or equivalently $m_{3/2}=m_{3/2}(C_0)$ in the top-down
approach.} Is this a true minimum or just an extremum? A {\em necessary}
condition is $\partial^2 V_1/\partial T^2_R>0$ at the minimum. We
obtain
\begin{equation}
T^2_R\,{\partial^2 V_1\over\partial T^2_R}=(p+3p^2)X{\partial V_1\over
\partial X}-2p^2V_1+{p^2\over64\pi^2}{\rm Str}{\cal M}^4,
\end{equation}
and at the minimum
\begin{equation}
\left(T_R^2\,{\partial^2 V_1\over\partial T_R^2}\right)_{\rm
minimum}={p^2\over32\pi^2}{\rm Str}{\cal M}^4.
\end{equation}
Therefore, if ${\rm Str}{\cal M}^4>0$, then the minimum in the moduli direction
is stable, and the vacuum energy is negative.

The necessary condition  for stability of the no-scale mechanism (${\rm
Str}{\cal M}^4>0$) has an important consequence: it imposes an upper bound on
$m_{3/2}$ as a function of the usual soft-supersymmetry-breaking parameters.
If we define the ratios
\begin{equation}
\xi_{3/2}\equiv{m_{3/2}\over m_{1/2}}\,,\quad
\xi_{0}\equiv{m_{0}\over m_{1/2}}\,,\quad
\xi_{A}\equiv{A\over m_{1/2}}\ ,
\end{equation}
then the upper bound is on the parameter $\xi_{3/2}$. To get a rough estimate
of this bound, we consider only the dominant contributions to ${\rm Str}{\cal
M}^4$, namely those from the three generations of (degenerate) squarks and
the gravitino:
\begin{equation}
{\rm Str}{\cal M}^4\approx 6\times12m^4_{\tilde q}-4m^4_{3/2}>0.
\end{equation}
Therefore
\begin{equation}
{m_{3/2}\over m_{\tilde q}}<(18)^{1/4}\approx2.06\ .
\label{m3/2bound}
\end{equation}
In practice we expect the factor to be slightly higher because of the
neglected contributions to ${\rm Str}{\cal M}^4$. Since to good approximation
one can write $m_{\tilde q}\approx m_0\sqrt{1+c_{\tilde q}/\xi^2_0}$, with
$c_{\tilde q}\sim4-6$ an RGE-dependent constant, we also have
\begin{equation}
{\xi_{3/2}\over \xi_0}\lsim2.06\sqrt{1+c_{\tilde q}/\xi^2_0}\ .
\label{eq:xi3/2bound}
\end{equation}

In addition to this necessary requirement for stability of the potential, the
absence of negative mass eigenstates in the neutral scalar sector provides more
stringent although model-dependent constraints which are analyzed in the next
section.

\section{New phenomenology: the neutral scalar sector}
Turning to the Higgs mass spectrum, interesting new possibilities beyond the
Supersymmetric Standard Model exist because of the new neutral scalar degrees
of freedom corresponding to the flat direction(s). Although the analysis of the
previous section was independent of the particular relation between $m_{3/2}$
and the flat direction(s), this relation enters the mass matrix of the real
neutral scalars. For simplicity, we analyze these new possibilities assuming
only one flat direction, but we believe the qualitative form of these results
persist in models with more flat directions. The resulting phenomenology neatly
divides into three cases, depending on the ratio $\gamma\equiv m_{3/2}/T_R$,
and each case has interesting and potentially observable new phenomena.
In the limit of exact CP conservation, the neutral real and imaginary degrees
of freedom decouple. Since the potential is independent of $T_I={\rm Im}(T)$,
this field is exactly massless and completely decouples, leaving the spectrum
of imaginary Higgs degrees of freedom unchanged. However, $T_R$ must be
included in a $3\times3$ mass matrix along with the usual two real Higgs
degrees of freedom.

Using the relation between $m_{3/2}$ and one flat direction in
Eq.~(\ref{eq:m3/2T}), and evaluating the derivatives with respect to $T_R$,
gives the following $3\times3$ mass matrix for the neutral real scalar degrees
of freedom
\begin{equation}
{1\over2}
\left(
\begin{array}{ccc}
{\partial^2 V_1/\partial h_1\partial h_1}&
{\partial^2 V_1/\partial h_1\partial h_2}&
{p\over64\pi^2}{1\over T_R}{\partial{\rm Str}{\cal M}^4/\partial h_1}\\
{\partial^2 V_1/\partial h_2\partial h_1}&
{\partial^2 V_1/\partial h_2\partial h_2}&
{p\over64\pi^2}{1\over T_R}\partial{\rm Str}{\cal M}^4/\partial h_2\\
{p\over64\pi^2}{1\over T_R}{\partial{\rm Str}{\cal M}^4/\partial h_1}
&{p\over64\pi^2}{1\over T_R}{\partial{\rm Str}{\cal M}^4/\partial h_2}
&{2p^2\over64\pi^2}{1\over T^2_R}{\rm Str}{\cal M}^4
\end{array}
\right).
\end{equation}
The analysis of this matrix is simplified by making a change of basis on the
Higgs fields to introduce some zero entries
\begin{equation}
\left(\begin{array}{ccc}a&b&0\\ b&c&d\\ 0&d&e\end{array}\right),
\end{equation}
where $a,b,c={\cal O}(m^2_{3/2})$, $d={\cal O}(\gamma m^2_{3/2})$, and $e={\cal
O}(\gamma^2 m^2_{3/2})$. In the limits $\gamma\gg1$ or $\gamma\ll1$, the form
of the eigenvectors can be identified as
\begin{equation}
\left(\begin{array}{c}\cos\phi\\ \sin\phi\\ \epsilon_1\end{array}\right),\qquad
\left(\begin{array}{c}-\sin\phi\\ \cos\phi\\
\epsilon_2\end{array}\right),\qquad
\left(\begin{array}{c}\epsilon_3\\ \epsilon_4\\ 1\end{array}\right),
\end{equation}
where $|\epsilon_{1,2,3,4}|\ll1$. In these two limits there is little mixing
between the Higgs and $T_R$ fields.

In the $\gamma\ll1$ limit we obtain the usual Higgs-boson masses with small
${\cal O}(\gamma^2m^2_{3/2})$ corrections to the squared masses. To first order
we obtain $m^2_{T_R}=e-ad^2/(ca-b^2)={\cal O}(\gamma^2m^2_{3/2})$, and there is
a new, very light scalar. Assuming that the usual $2\times2$ Higgs mass matrix
has a positive determinant, $m^2_{T_R}>0$ is equivalent to the whole mass
matrix having a positive determinant.
In the $\gamma\gg1$ limit the Higgs masses are given to first order by
\begin{equation}
m^2_{H_R}={1\over2}\left\{c+a-{d^2\over e}\pm\sqrt{\left({d^2\over
e}-c+a\right)^2+4b^2}\right\},
\end{equation}
which depart from their usual value in models without the $T_R$ field
because of the new $d^2/e$ terms. Again, assuming the usual $2\times2$ Higgs
mass matrix has a positive determinant, $m^2_{H_R}>0$ is equivalent to the
whole mass matrix having a positive determinant.  Note that the the resulting
Higgs masses are independent of $\gamma$, since $d^2/e={\cal O}(m^2_{3/2})$.
The mass of $T_R$ is $e\sim{\cal O}(\gamma^2m^2_{3/2})\gg m^2_{3/2}$. In the
case where $\gamma\sim1$, the mass eigenstates are general mixtures of the
Higgs and $T_R$ fields, and must be analyzed as a function of parameter space.

We reach the very interesting conclusion that no-scale supergravity always
results in phenomenology beyond its global supersymmetric counterpart because
of the fields associated with the flat direction.  The real neutral scalar
phenomenology falls into three cases. For $\gamma\ll1$, there is an additional
scalar with  $m^2\sim{\cal O}(\gamma^2m^2_{3/2})$ and unaltered Higgs
predictions. For $\gamma\gg1$ the usual predictions for the Higgs masses are
altered.  For $\gamma\sim1$, there is a rich new phenomenology arising from a
mass matrix significantly mixing the $H_R$ and $T_R$ fields which has detailed
dependence on the parameter space of the model. This `no-lose' situation offers
hope for experimentally verifying no-scale supergravity. In addition, the
distinct phenomenological properties of the three cases provides an
experimental handle on the parameter $\gamma$, and thereby information about
the K\"ahler potential.

\section{Numerical investigations}
The unknown parameter space of our model may be
taken as ($\tan\beta$,$m_t$,$m_{1/2}$,$\xi_0$,$\xi_A$,$\xi_{3/2}$).  Given
these parameters the Higgs mixing terms $\mu$ and $B$ can be calculated
from the usual minimization conditions \cite{aspects}.  The parameter
$C$ can be calculated from the requirement that the potential is
a minimum with respect to $m_{3/2}$.  Although this may be accomplished
by taking derivatives numerically, a more efficient (and accurate)
method results from setting Eqs.~(\ref{eq:V1}) and (\ref{eq:V1min}) equal and
solving for $C_{1/2}\equiv{C\xi_{3/2}^4}\ $.
\begin{equation}
C_{1/2}\,m_{1/2}^4=-\left[
V_0+{1\over64\pi^2}{\rm Str}{\cal M}^4\left(\ln{{\cal
M}^2\over Q^2}-{3\over2}\right)+{1\over128\pi^2}{\rm Str}{\cal M}^4
\right].
\end{equation}
The problem may be further simplified by separating out the contribution (in
the supertraces) of the gravitino to $C_{1/2}$ from the other contributions,
\begin{equation}
C_{1/2}=C_{1/2}^g+\widetilde C_{1/2}\ .
\end{equation}
Note that $\widetilde C_{1/2}$ depends on the five parameters
($\tan\beta$,$m_t$,$m_{1/2}$,$\xi_0$,$\xi_A$) and is independent of
$\xi_{3/2}$.  On the other hand, $C_{1/2}^g$ depends only on
($\xi_{3/2}$,$m_{1/2}$) and is independent of the other four parameters:
\begin{equation}
C_{1/2}^g={\xi^4_{3/2}\over16\pi^2}
\left(\ln{\xi_{3/2}^2m_{1/2}^2\over{M_Z^2}}-1\right).
\end{equation}

Next, consider the scaling of $C_{1/2}$ (from $M_Z$ to the unification scale
$M_U$) which obeys the same RGE as $C$, Eq.~(\ref{eq:etaRGE}),
with $m_{1/2}$ replacing $m_{3/2}$
\begin{equation}
C_{1/2}(M_U)=C_{1/2}(M_Z)+\Delta C_{1/2}\ .
\end{equation}
Since the RGE is linear, it can also be separated into a part due to the
contribution of the gravitino and a part due to all the other particles, and we
can write
\begin{equation}
\Delta C_{1/2}=\Delta C_{1/2}^g+\Delta\widetilde C_{1/2}\ ,
\end{equation}
with a simple expression for $\Delta C_{1/2}^g$
\begin{equation}
\Delta C_{1/2}^g={\xi_{3/2}^4\over8\pi^2}\ln{M_U\over{M_Z}}\ .
\end{equation}

\begin{figure}[p]
\vspace{6in}
\includegraphics{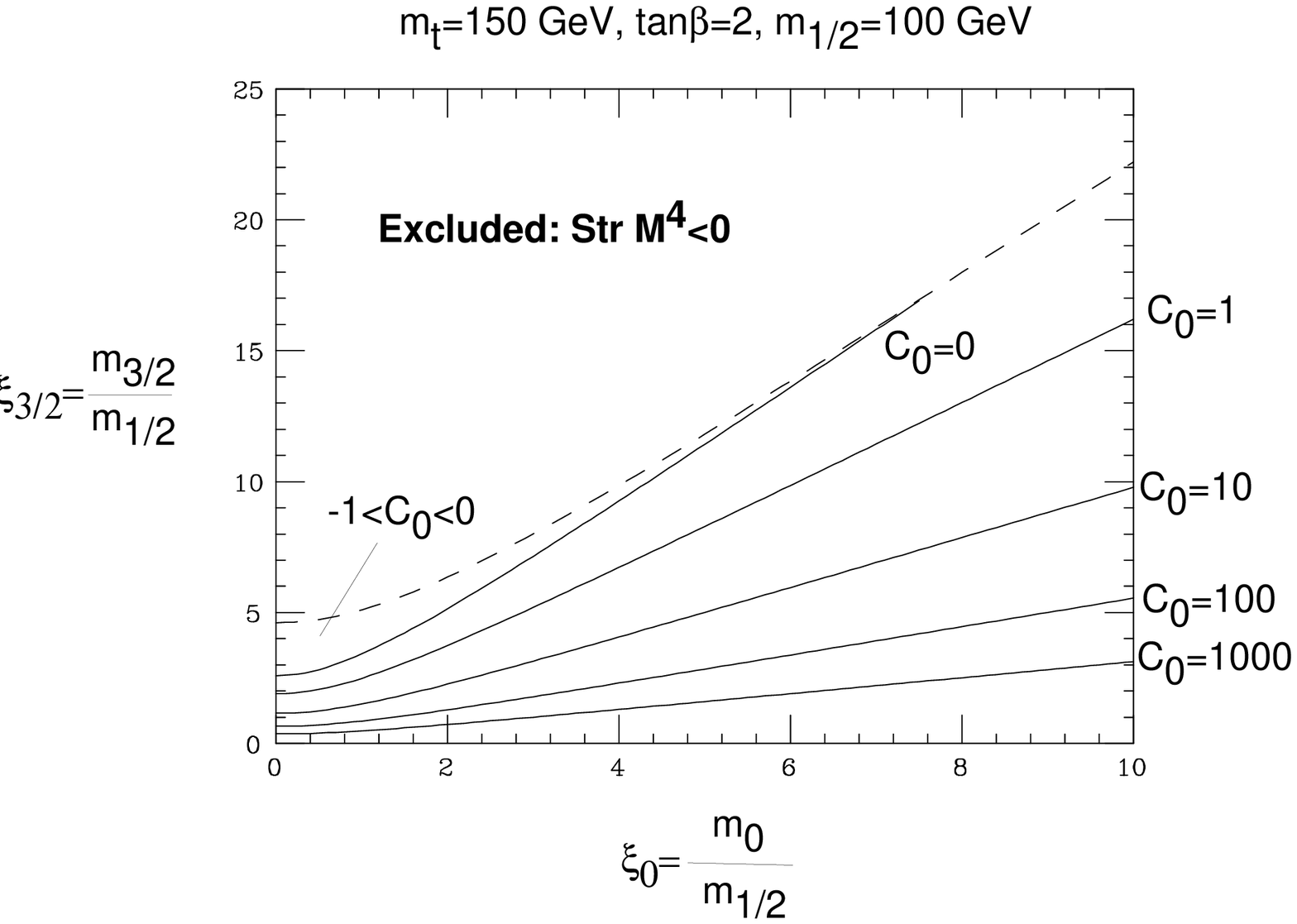}
\caption{Contours of constant values of $C_0=C(M_U)$ in the
$(\xi_0,\xi_{3/2}$) plane. The dependence on $m_{1/2}$ is very mild; $m_t$ and
$\tan\beta$ are also immaterial. The area above the dashed line is excluded by
the stability of the no-scale mechanism which requires ${\rm Str}{\cal
M}^4>0$.}
\label{Fig}
\end{figure}

We have given the explicit analytic expressions for $C_{1/2}^g$ and
$\Delta C_{1/2}^g$, whereas $\widetilde C_{1/2}$ and $\Delta\widetilde C_{1/2}$
must be calculated numerically as a function of the five-dimensional parameter
space ($\tan\beta$,$m_t$,$m_{1/2}$,$\xi_0$,$\xi_A$). However, within the
phenomenologically viable area of this five-dimensional parameter space,
we find the results nearly independent of $\tan\beta$, $m_t$, and $\xi_A$, with
a small dependence on $m_{1/2}$. Therefore, we present our final results for
$C_0=C(M_U)$ in the ($\xi_0$,$\xi_{3/2}$) plane shown in
Figure~\ref{Fig}. The particular plot shown is for $m_t=150\GeV$,
$\tan\beta=2$, $\xi_A=0$, and $m_{1/2}=100\GeV$ although changing these
variables shifts the contours only slightly.

Note that the stability of the potential puts an upper bound on $\xi_{3/2}$
which is well approximated by Eq.~(\ref{eq:xi3/2bound}).  If there were some
upper bound on the magnitude of $C_0$ from naturalness or string
considerations, this would give a lower bound on $\xi_{3/2}$.  For example, if
we require $|C_0|<10$, then for $\xi_0=0$, $1.2<\xi_{3/2}<4.6$.  For
$\xi_0=4$ the bound becomes $4.1<\xi_{3/2}<9.9$. Turning things around, typical
supersymmetry breaking scenarios entail $m_0=m_{3/2}$ (\ie,
$\xi_{3/2}=\xi_0$) \cite{IL} or $m_{1/2}=m_{3/2}$ (\ie, $\xi_{3/2}=1$)
\cite{A}. In both cases Fig.~\ref{Fig} shows that $C_0>10$ is required.
These types of bounds are very interesting in the context of explicit
predictions for the various $\xi's$ in string models, and will be even more
interesting in the context of sparticle spectroscopy \cite{FHKN}.

\section{Simple models}
\label{sec:simple}
Assuming that the only fundamental scales in the theory are
$M_{Pl}$ and $m_{3/2}$, dimensional arguments for the simplest
relation between $T_R$ and $m_{3/2}$ give
\begin{equation}
m_{3/2}^2={\alpha\kappa^{-p-2}\over(T_R)^p}\ ,
\label{eq:25}
\end{equation}
where $\kappa=\sqrt{8\pi}/M_{Pl}$ is the gravitational coupling. This relation
is the analogue of Eq.~(\ref{eq:m3/2T}) for $\Lambda\sim M_{Pl}$. From this
relation we obtain
\begin{equation}
\gamma={m_{3/2}\over T_R}\sim{1\over\alpha^{1/p}} \left({m_{3/2}\over
M_{Pl}}\right)^{1+2/p}.
\label{eq:26}
\end{equation}
Thus, the three scenarios for $\gamma$ are related to the functional
dependence of the gravitino mass: $\gamma\gg1$ implies
$\alpha\ll(m_{3/2}/M_{Pl})^{p+2}$,
$\gamma\ll1$ implies $\alpha\gg(m_{3/2}/M_{Pl})^{p+2}$, and
$\gamma\sim1$ implies $\alpha\sim(m_{3/2}/M_{Pl})^{p+2}$.

We now consider a choice for the K\"ahler function which possesses the three
properties described in the introduction. In string models built
within the free-fermionic formulation, one can show that the untwisted sector
fields split up into three sets, each with a separate contribution to the
K\"ahler function. In the simplest models these contributions can be written as
\cite{FKPZII,LNY94}
\begin{equation}
G=-\ln(S+\bar S)-\sum_{A=1}^3\ln Y_A+K_{\rm TS}+\ln|W|^2,
\end{equation}
with
\begin{equation}
Y_A=1-\sum_{i_A}\alpha_{i_A}\bar\alpha_{i_A}
+\coeff{1}{4}\sum_{i_A}(\alpha_{i_A}\alpha_{i_A})
(\bar\alpha_{i_A}\bar\alpha_{i_A}),
\end{equation}
$K_{\rm TS}$ the twisted sector contribution, and $W$ the superpotential. To
make things simple we will neglect the second and third sets (\ie, set
$\alpha_{i_{2,3}}=0$) as well as the twisted sector contribution. Also, we will
only consider three fields in the first set. This approximation should suffice
for our present purposes. Through an analytic field redefinition \cite{FKPZII},
our simplified K\"ahler function reduces to
\begin{equation}
G=-\ln(S+\bar S)-\ln[\a\b-\c^2]+\ln|W(\phi)|^2,
\end{equation}
where $T$ and $U$ are the moduli fields, and $\phi$ is a charged matter field.
The scalar potential is then
\begin{equation}
V=e^G\left\{-\c(\h+\hb)+\coeff{1}{2}\left[\a\b+\c^2\right]|\h|^2\right\},
\end{equation}
which is not obviously positive semi-definite. Considering minima which
preserve the gauge symmetry, \ie, with $\VEV{\phi}=\VEV{\bar\phi}=0$, the
potential reduces to
\begin{equation}
V_{\VEV{\phi}=\VEV{\bar\phi}=0}={1\over2}{|W|^2\over(S+\bar S)}\,|\h|^2,
\end{equation}
which is positive semi-definite. Therefore the minima, which occur for
$\VEV{\h}=0$, have zero vacuum energy. Moreover, the potential is manifestly
$T$- and $U$-independent, \ie, we have two flat directions.\footnote{The
value of $S$ is also undetermined. This vev could be fixed by giving $W$
and $S$-dependence.} The gravitino mass is given by
\begin{equation}
m^2_{3/2}={\VEV{|W|^2}\over\VEV{(S+\bar
S)\a\b}}={g^2\VEV{|W|^2}\over\VEV{\a\b}}\ ,
\end{equation}
and is undetermined because of the $T$ and $U$ flat directions. The goldstino
is a to-be-determined linear combination of $\widetilde S$, $\widetilde T$,
and $\widetilde U$. Whether the third property (\ie, ${\rm Str}{\cal M}^2=0$)
is satisfied or not depends on the mechanism for supersymmetry breaking
\cite{FKZ}. In tree-level breaking via coordinate-dependent compactifications
one has $W=w+\cdots$, with the limit $W\to w$ for $\phi\to0$, and $w$ a
constant of order 1. The  ${\rm Str}{\cal M}^2=0$ condition then imposes a
non-trivial constraint on the number of fields belonging to each of the three
sets of the theory \cite{FKZ}. Writing $m^2_{3/2}=g^2w^2/R^2$, with
$R^2=\a\b\kappa^4$ the ``radius" of the compactified dimension, it is clear
that $m_{3/2}\sim1\TeV$ requires $R^{-1}\sim1\TeV$. In this
``decompactification" limit there is a tower of Kaluza-Klein string massive
states which can be as light as a few TeV and have distinct experimental
signatures \cite{A}.

Let us take $T=U$, which then gives $p=2$ and $\alpha=g|W|$ in
Eq.~(\ref{eq:25}), and therefore $\gamma\sim\alpha^{-1/2}\,(m_{3/2}/M_{Pl})^2$.
If $\alpha\sim1$, as expected in coordinate-dependent compactifications, then
$\gamma\ll1$. On the other hand, if $\alpha\ll1$, as expected in gaugino
condensation models, the $\gamma\sim1$ may be possible. More possibilities may
exist in more general supersymmetry breaking scenarios.

\section{Conclusions}
The solution to the gauge hierarchy problem via the no-scale mechanism selects
a special class of string-derived supergravities. We have studied some simple
examples of these very appealing models and have shown that the no-scale
mechanism leads to a stable minimum of the electroweak potential for large
regions of the soft-supersymmetry-breaking parameter space. Moreover, this
stability requirement can be neatly encoded in the upper bound $m_{3/2}\lsim
2m_{\tilde q}$. The no-scale mechanism also entails mixing of the Higgs
fields with the moduli, which could be experimentally observable in some
supersymmetry breaking scenarios. These two results appear to be {\em unique}
to the no-scale mechanism and, if verified experimentally, would constitute the
``smoking gun" of no-scale supergravity. We are in the process of addressing
the laboratory and cosmological implications of such no-scale supergravity
properties. We have also studied the dependence on the
soft-supersymmetry-breaking parameters of the residual vacuum energy at the
unification scale ($C_0\,m^4_{3/2}$). We find that in typical models one must
require $C_0>10$. Our results should be useful to string model builders
searching for string no-scale supergravity models with phenomenologically
viable values of $C_0$.

\section*{Acknowledgments}
This work has been supported in part by DOE grant DE-FG05-91-ER-40633.

\newpage

\clearpage

\end{document}